\documentclass[12pt,preprint]{aastex}

\usepackage{graphicx}

\slugcomment{Not to appear in Nonlearned J., 45.}

\shorttitle{Yaglom law in the solar wind}
\shortauthors{Gogoberidze et al.}

\begin{document}

\title{Yaglom law in the expanding solar wind}

\author{G. Gogoberidze}
\affil{Dipartimento di Fisica, Universit\`a della Calabria, I-87036 Rende, Italy}
\affil{Institute of Theoretical Physics, Ilia State University, 3/5 Cholokashvili ave., 0162 Tbilisi, Georgia}
\email{g.gogoberidze@warwick.ac.uk}

\author{S. Perri and V. Carbone}
\affil{Dipartimento di Fisica, Universit\`a della Calabria, I-87036 Rende, Italy}

\begin{abstract}
We study the Yaglom law, which relates the mixed third order structure function to the average dissipation rate of turbulence, in a uniformly expanding solar wind by using the two scales expansion model of magnetohydrodynamic (MHD) turbulence. We show that due to the expansion of the solar wind two new terms appear in the Yaglom law. The first term is related to the decay of the turbulent energy by nonlinear interactions, whereas the second term is related to the non-zero cross-correlation of the Els\"asser fields. Using magnetic field and plasma data from WIND and Helios 2 spacecrafts, we show that at lower frequencies in the inertial range of MHD turbulence the new terms become comparable to Yaglom's third order mixed moment, and therefore they cannot be neglected in the evaluation of the energy cascade rate in the solar wind. 
\end{abstract}


\section{Introduction}

The standard theory of the solar wind \citep{P63} predicts the existence of a plasma flow, coming from the hot solar corona, that expands radially in the interplanetary medium without being further heated. In the case of a spherically symmetric adiabatic expansion, the proton temperature $T_p$ should vary as $T_p\sim R^{-4/3}$ with the distance from the sun $R$. However, observations from the Helios spacecraft from $0.3$ to $1$ AU showed a temperature radial profile $T_p\sim R^{-0.9}$  \citep{TEA95}, thus implying {\it in-situ} heating of the solar wind plasma during the expansion (see \citet{Marsch82,Tu95} and references therein). A first possible scenario for solar wind heating was proposed by \citet{Coleman68}, who ascribed that phenomenon to the existence of a turbulent cascade, both in the magnetic and in the velocity fields, that is eventually dissipated producing internal energy. This has been considered an efficient mechanism at least for $R<10 ~ {\rm AU}$. This idea has been supported by the fact that the proton temperature is positively correlated with the amplitude of magnetic field and velocity fluctuations \citep{BD71}. 
Kolmogorov's theory for stationary, isotropic, and fully developed fluid turbulence \citep{K41} predicts for the one dimensional spectral energy density $E(k)$ in the inertial interval
\begin{equation}
E(k)=C_K \varepsilon^{2/3}k^{-5/3}, \label{eq:KOL}
\end{equation}
where $C_K\approx 1.6$ is the Kolmogorov constant and $k$ is the wave number. A similar approach has also been extended to magnetized fluids, as the solar wind \citep{Iroshnikov63,Kraichnan65,DMV80,GS95}. A recent study performed by \citet{MSF08} showed that the energy cascade rates derived by this method are not accurate enough and poorly agree with the proton heating rates.

Assuming a model for spherical symmetric expansion of the solar wind \citep{Verma95} and using proton temperature data, \citet{VEA07} found the following expression for the proton heating rate per unit mass at $1$ AU
\begin{equation}
\varepsilon_p=3.6\times 10^{-5} T_p V_{sw} ~{\rm J/(Kg \cdot s)}, \label{eq:PROTON}
\end{equation}
where $V_{sw}$ is the solar wind velocity in ${\rm km/s}$ and the proton temperature $T_p$ is measured in $K$.
Analysis of the solar wind data performed by \citet{PEA90} and theoretical predictions by \citet{LEA99} suggested that {\it in-situ} electron heating rate is comparable to $\varepsilon_p$. Equation (\ref{eq:PROTON}) implies that for fast warm streams in the solar wind the heating rate is of the order $\varepsilon_p\sim 10^4~{\rm J/(Kg \cdot s)} $, whereas for relatively slow, cool streams a typical value of the heating rate is $\varepsilon_p\sim ~10^3 {\rm J/(Kg\cdot s)}$.

An alternative method for the derivation of the energy cascade rate is to use the Yaglom law, which relates the third order structure function to the mean energy cascade rate and represents one of the most fundamental relations in the theory of turbulence \citep{Yaglom49,C67,Frisch95}. Generalization of the Yaglom law for MHD turbulence was derived by \citet{Politano95,PP98}. \citet{SEA07}, \citet{CEA08}, and \citet{MSF08} used solar wind data from various satellites to derive the energy cascade rate by means of the Yaglom law for isotropic MHD turbulence. The latter showed that the energy cascade rates obtained through the Yaglom law are in qualitative agreement with the predictions of equation (\ref{eq:PROTON}). Recently \citet{OEA11} used multi spacecraft data from the Cluster mission to study the energy cascade rate via the anisotropic form of the Yaglom law.
It should also be noted that relatively high level of measurement uncertainties in the plasma data have much less influence on Yaglom's third order moment than on the second order moment of MHD turbulence \citep{GCHD12}.

In all those previous studies homogeneity of turbulence has been assumed, while solar wind expansion has not been taken into account. In this Letter we study the influence of the solar wind expansion on the Yaglom law. Namely, using the two scale expansion model developed by \citet{ZM90}, where the `global' variable related to the solar wind expansion is considered as `slow' variable, whereas nonlinear interactions are mainly determined by local, `fast', variables, we derive a modified expression for the Yaglom law. We show that two extra terms appear in the Yaglom law due to the solar wind expansion. The first extra term is related to non-WKB decay of the turbulent energy due to nonlinear interactions, whereas the second term is caused by the interaction between large-scale fields and the cross correlation of the small scale inward and outward propagating Alfv\'en waves. Using data from WIND and Helios 2 spacecrafts, we show that the novel terms in the expansion modified Yaglom law become comparable to Yaglom's third order moment at larger time scales in the inertial range of solar wind turbulence and, therefore, make significant contribution to the estimate of the energy dissipation rate.

The Letter is organized as follows: modified Yaglom law in the uniformly expanding environment is derived in Sec. 2. The solar wind data analysis is presented in Sec. 3. Conclusions are given in Sec. 4.

\section{Theoretical analysis}

We consider incompressible MHD turbulence in the presence of a constant magnetic field ${\bf B}_0$. The Els\"asser variables ${\bf Z}^\pm={\bf v}\pm{\bf b}/\sqrt{4\pi \rho}$, the eigenfunctions of counter propagating Alfv\'en waves, are usually considered as the most fundamental variables to study MHD turbulence \citep{B}. 
The dynamics of the Els\"asser variables is governed by the incompressible MHD equations
\begin{equation}
\left( \frac{\partial}{\partial t} \mp {\bf V}_A \cdot {\bf \nabla} \right) {\bf Z}^\pm + ({\bf Z}^\mp \cdot {\bf \nabla}){\bf Z}^\pm +{\bf \nabla} p+\lambda^\pm \nabla^2{\bf Z}^+ + \lambda^\mp \nabla^2{\bf Z}^-=0. \label{eq:01}
\end{equation}
Here $p$ is the total (hydrodynamic plus magnetic) pressure, ${\bf V}_A \equiv {\bf B_0}/\sqrt{4\pi \rho}$ is the Alfv\'en velocity related to the background magnetic field, $\rho$ is the mass density, $\lambda^\pm=(\nu\pm \mu)/2$ where $\nu$ is the kinematic viscosity and $\mu$ is the magnetic diffusivity. Although the solar wind plasma is weakly collisional and, consequently, turbulent fluctuations are mainly damped by kinetic mechanisms, in our further analysis we will include the collisional dissipation terms in equation (\ref{eq:01}) similarly to the other studies \citep{PP98,CSM09}. Since Yaglom's law is observed at relatively small scales in the inertial range of the solar wind turbulence, the precise form of the dissipation mechanism, acting at very small scales, seems to be unessential for the present study.

It is well known that smooth average properties of the solar wind vary on length scales of the order of the heliospheric distance $R$. On the other hand, turbulent fluctuations have a correlation length that is much smaller than $R$ \citep{Matthaeus05}. This scale separation allows for formulation of the so-called WKB-like transport equations for MHD turbulence in the solar wind \citep{TPW84,ZM90}. 
In the framework of this approach the magnetic field and the velocity are considered as a sum of mean and fluctuating parts. The mean parts depend only on global, slowly varying coordinate (denoted by ${\bf R}$ hereafter) and is related to the large scale inhomogeneities in the solar wind plasma, whereas fluctuating (turbulent) fields depend on both the slow and a rapidly varying coordinate, denoted here by ${\bf r}$. 
Large and small scales can be separated, assuming ergodicity \citep{Matthaeus82}, by introducing an ensemble averaging operation $\langle \rangle$. 
Here we use the two scales expansion model presented in \citet{ZM90}: equations for fluctuations in the expanding environment are
\begin{eqnarray}
\frac{\partial {\bf z}^\pm}{\partial t} + \left( {\bf U} \mp {\bf V}_A \right) \cdot {\bf \nabla}  {\bf z}^\pm + \frac{{\bf z}^\pm - {\bf z}^\mp}{2} \nabla \cdot \left( \frac{{\bf U}}{2} \pm {\bf V}_A \right) \nonumber \\ + {\bf z}^\pm \cdot \left( \nabla {\bf U} \pm \frac{1}{\sqrt{4\pi \rho}}\nabla {\bf B}_0 \right) +({\bf z}^\mp \cdot {\bf \nabla}){\bf z}^\pm = \nonumber \\ -\frac{1}{\rho} {\bf \nabla} p+\lambda^\pm \nabla^2 {\bf z}^+ + \lambda^\mp \nabla^2 {\bf z}^-. \label{eq:WKB}
\end{eqnarray}
Here ${\bf z}^\pm$ denote the turbulent fluctuations of the Els\"asser variables and ${\bf B}_0$ is the large scale mean magnetic field. This equation is derived assuming incompressibility  with respect to the small scale variable ${\bf r}$, i.e. ${\bf \nabla}_{\bf r}\cdot {\bf v}=\nabla_{\bf r} \cdot {\rho}=0$.

In hydrodynamics a mean or large-scale flow sweeps the small-scale eddies without affecting the energy
transfer between length scales. ͑In magnetohydrodynamics a mean (or large-scale) magnetic field ${\bf B}_0$ sweeps oppositely propagating fluctuations ${\bf z}^-$ and ${\bf z}^+$, which affects the energy transfer. Therefore, contrary to hydrodynamics, in case of MHD turbulence both linear (sweeping) and nonliear (streining) timescales are important for the dynamics of turbulence in the inertial interval (for a review see \citet{ZMD04,Z10}). The consept of the scale separation introduced above implies that the energy cascade timescale (which in general case depends on both the sweeping and straining of turbulent fluctuations) is much less compared to the characteristic timescale of the solar wind expansion $R/U$.

 For further simplifications of the analysis we make several additional assumptions: we consider a uniform radial expansion, i.e., ${\bf U}=(U,0,0)$ with $U=const$, and a constant large scale mean magnetic field ${\bf B}_0$; we also assume $U\gg V_A$, which is reasonable in the case of the solar wind (being $U\sim 400$ km/s and $V_A\sim 40$ km/s), and finally we assume $\lambda^+=\lambda^-=\nu$. For a uniform expansion $\nabla {\bf U}=2U/R$, so that equation (\ref{eq:WKB}) reduces to
\begin{eqnarray}
\frac{\partial {\bf z}^\pm}{\partial t} + {\bf U}\cdot {\bf \nabla} {\bf z}^\pm \mp {\bf V}_A \cdot {\bf \nabla} {\bf z}^\pm + ({\bf z}^\mp \cdot {\bf \nabla}){\bf z}^\pm  +({\bf z}^\pm - {\bf z}^\mp )\frac{U}{2R} = -\frac{1}{\rho} {\bf \nabla} p+\nu \nabla^2 ({\bf z}^+ + {\bf z}^-). \label{eq:WKB1}
\end{eqnarray}
As we see, due to the expansion, two extra terms appear in equation \ref{eq:WKB1} compared to the standard incompressible MHD equation for the turbulent fluctuations \citep{Bruno05}. The second term on the LHS of equation (\ref{eq:WKB1}) describes convective acceleration related to the solar wind expansion, whereas the fifth term describes interaction between large scale fields and the cross correlation of the Els\"asser fields \citep{ZM89}. 

The derivation of the Yaglom law in MHD turbulence \citep{PP98} has already been presented in details without the expansion effects by \citet{CSM09}. 
Derivation of the Yaglom law implies the following steps: one should consider equation (\ref{eq:WKB1}) in two different points, ${\bf x}^\prime$ and  ${\bf x}$, thus allowing to define ${\bf R}=({\bf x}^\prime+{\bf x})/2$ and ${\bf r}={\bf x}^\prime-{\bf x}$. Subtracting those equations in ${\bf x}^\prime$ and  ${\bf x}$ we can derive dynamical equations for the differences of the Els\"asser fields $\Delta {\bf z}^\pm={\bf z}^{\pm}({\bf x}^\prime)-{\bf z}^\pm({\bf x})$. We then multiply the obtained equations by $\Delta {\bf z}^\pm$, we ensemble average and take the trace. Under the assumption of local homogeneity, one finally ends up with the dynamical equation for the evolution of $\langle |\Delta z^\pm|^2 \rangle$ (equation (12) in \cite{CSM09}). It can be readily shown that the two extra terms related to the expansion evolve as follows: the third term on the LHS of equation (\ref{eq:WKB1}) gives $U\partial_R \langle |\Delta z^\pm|^2 \rangle$, whereas the forth term gives $U\langle |\Delta z^\pm|^2 \rangle/R - U\langle \Delta {\bf z^+}\cdot \Delta{\bf z^-}\rangle/R$. Therefore, the dynamical equation for $\langle |\Delta z^\pm|^2 \rangle$ takes the form
\begin{equation}
\frac{\partial }{\partial t} \langle |\Delta z^\pm|^2 \rangle  +\nabla \cdot \langle \Delta {\bf z}^\mp |\Delta z^\pm|^2 \rangle + \frac{U}{R} \frac{\partial }{\partial R}  (R \langle |\Delta z^\pm|^2 \rangle )- \frac{U}{R} \langle \Delta {\bf z}^+ \cdot \Delta {\bf z}^- \rangle = 2\nu \nabla^2\langle |\Delta z^\pm|^2 \rangle -\frac{4}{3}\nabla \cdot (\varepsilon^\pm {\bf r}). \label{eq:ANI}
\end{equation}
Here $\varepsilon^\pm \equiv \nu \langle \partial_i z_j^\pm \partial_i z_j^\pm \rangle$ (where indices indicate the summation over the vector components) are the pseudo energy dissipation rates of the corresponding Els\"asser fields. Equation (\ref{eq:ANI}) is valid even for anisotropic MHD turbulence \citep{CSM09}. A further simplification implies assumption of local isotropy. Although MHD turbulence is known to be anisotropic in the inertial range due to the influence of the mean magnetic field (see, e.g., \cite{B}), study performed by \citet{SEA09} showed that the analysis of solar wind data based on isotropic and anisotropic versions of the Yaglom law gives very similar results for the energy dissipation rate of MHD turbulence, therefore the assumption of isotropy is reasonable for the purposes of the presented study. In this framework the gradient and the Laplacian operators become 
\begin{equation}
\nabla=\frac{2}{r}+\frac{\partial}{\partial r},~~~\nabla^2=\left(\frac{2}{r}+\frac{\partial}{\partial r}\right) \frac{\partial}{\partial r}. \label{eq:ISO}
\end{equation}
Considering stationary turbulence (thus dropping the first term on the LHS of equation (\ref{eq:ANI})), which could be a reasonable approximation in certain fast wind streams, the integration of equation (\ref{eq:ANI}) in the inertial range (where the influence of the dissipation term on the RHS can be neglected) yields
\begin{equation}
 \langle \Delta {z}^\mp_r |\Delta z^\pm|^2 \rangle 
+\frac{U}{Rr^2} \int_0^r y^2 \frac{\partial }{\partial R}  (R \langle |\Delta z^\pm|^2 \rangle ){\rm d}y-  \frac{U}{Rr^2} \int_0^r y^2 \langle \Delta {\bf z}^+ \cdot \Delta {\bf z}^- \rangle {\rm d}y= -\frac{4}{3} \varepsilon^\pm r. \label{eq:YAGr}
\end{equation}
This equation represents a generalization of the Yaglom law for uniformly expanding solar wind. 
The second term on the LHS of equation (\ref{eq:YAGr}) is related to the nonlinear interactions of the fluctuations. Indeed, as it was shown by \cite{MT90}, in the absence of nonlinear interactions and cross correlations, the energy densities of counter propagating Alfv\'en waves would have a WKB dependence on heliospheric distance $E^\pm(R)\rho(R)^{-1/2}(U^2\mp V_A^2)=const$ \citep{Marsch89}. In the considered case of the uniform expansion ($U=const$, $\rho(R)\sim 1/R^2$) and for $U\gg V_A$, it is easy to show that for fluctuations following WKB scaling, namely $E^\pm(R)\sim 1/R$, the second term on the LHS in equation (\ref{eq:YAGr}) vanishes. Therefore, this term is directly related to the decay of the turbulent energy by means of nonlinear interactions during radial expansion. Note that a similar term also appears in the modified Yaglom law for decaying hydrodynamic turbulence \citep{DAZA99}, which is known to dominate over the Yaglom's third order term for relatively large separation $r$. 

The second extra term related to the solar wind expansion (the third term on the LHS of equation (\ref{eq:YAGr})) is due to the cross correlation of the Els\"asser fields. The identity $\langle \Delta {\bf z}^+ \cdot \Delta {\bf z}^- \rangle=\langle \Delta {v}^2 \rangle - \langle \Delta {b}^2 \rangle/4\pi\rho$ shows that the cross correlation is proportional to the so-called residual energy, i.e., the difference between kinetic and magnetic energies. Taking into account that both solar wind data and numerical simulations of MHD turbulence show an excess of magnetic energy with respect to the energy contained in the velocity fluctuations at all scales in the inertial range (see, e.g., \citet{Tu91,Bavassano98,GCH12}), we conclude that the third term on the LHS of equation (\ref{eq:YAGr}) is positive.

Finally, it is worth noting that a dimensional analysis of equation (\ref{eq:YAGr}) yields for a ratio between the Yaglom third order term and the terms related to the expansion to be of order $(v_{rms}/L)/(U/R)$, where $v_{rms}$ is the rms value of the velocity fluctuations. At $1~{\rm AU}$ this ratio has the order $\gtrsim 10$ and one can conclude that expansion effects are small (or even negligible) in the inertial range of the solar wind turbulence. As we show below, this is not the case and at least for larger time scales in the inertial range these terms are of the same order of magnitude. The reason is that the Yaglom law contains ensemble average of third order quantity which does not have specific sign, and consequently, the result obtained after averaging is significantly reduced compared to the dimensional estimate $\langle \Delta {z}^\mp_r |\Delta z^\pm|^2 \rangle\sim v_{rms}^3$. 

To study the relative importance of the third order and of the solar wind expansion terms, we performed data analysis of the data from WIND and Helios 2 satellites. 
The decay term (second term on the LHS) contains derivative with respect to the heliospheric distance, so that it requires data of turbulent fluctuations at least at two different heliospheric distances. Due to this reason, first we use 3 sec resolution plasma and magnetic field data from the WIND satellite to compare the third order Yaglom and the cross correlation terms. Then we use 81 sec resolution data from Helios 2 collected from the same corotating plasma stream at different heliospheric distances \citep{MT90,Tu95} to assess the possible importance of the decay term. 

\section{Solar Wind Data Analysis}

In order to estimate Yaglom's third order term and the cross correlation term, we use magnetic field data from the Magnetic Field Investigation (MFI) instrument on board WIND at $3$ sec resolution \citep{LEP95}. Density and velocity data are provided by the three dimensional plasma (3DP) instrument \citep{LIN95}. 
We use observations made during a quiet fast stream. The start time of the interval is 10:00 of 2008 February 02 and the end time is 00:00 of
2008 February 04. During this interval, the solar wind speed remained
above $550~{\rm km/s}$. The energy of compressive fluctuations was an order of magnitude lower than that of incompressible fluctuations and, consequently, magnetic and velocity fluctuations, being mainly
Alfv\'enic, were dominated by the components perpendicular to the local mean field. 
We made use of Taylor's hypothesis \citep{T38} $r=\pm U\tau$ to relate temporal changes $\tau$ in the observational data to spatial variations $r$ of the turbulent fields, by using the solar wind speed $U$ as a transformation parameter. In the last expression the plus sign corresponds to the case when the radial axis is directed toward the sun, for instance in Geocentric Solar Ecliptic (GSE) coordinates (where the x axis points toward the sun and the z axis is perpendicular to the plane of the earth's orbit around the sun), the minus sign corresponds to the case when the radial axis is directed away from the sun, as in the Radial-Tangential-Normal (RTN) coordinate system. Using Taylor's hypothesis, equation (\ref{eq:YAGr}) can be rewritten as
\begin{equation}
\alpha Y^\pm + D^\pm + M = -\varepsilon^\pm, \label{eq:YAGt}
\end{equation}
where $Y^\pm= 3\langle \Delta {z}^\mp_r |\Delta z^\pm|^2 \rangle/4Ut$ is the Yaglom term, $D^\pm={3U} \int_0^t y^2 \frac{\partial }{\partial R}  (R \langle |\Delta z^\pm|^2 \rangle ){\rm d}y/{4Rt^3}$ is the decay term, $M= - {3U} \int_0^t y^2 \langle \Delta {\bf z}^+ \cdot \Delta {\bf z}^- \rangle {\rm d}y/{4Rt^3} $ is the cross correlation term, $\alpha$ is $+1$ if $r=U\tau$ and $-1$ for $r=-U\tau$, depending on the reference frame, as discussed above. 

\begin{figure}
\epsscale{0.99}
\plotone{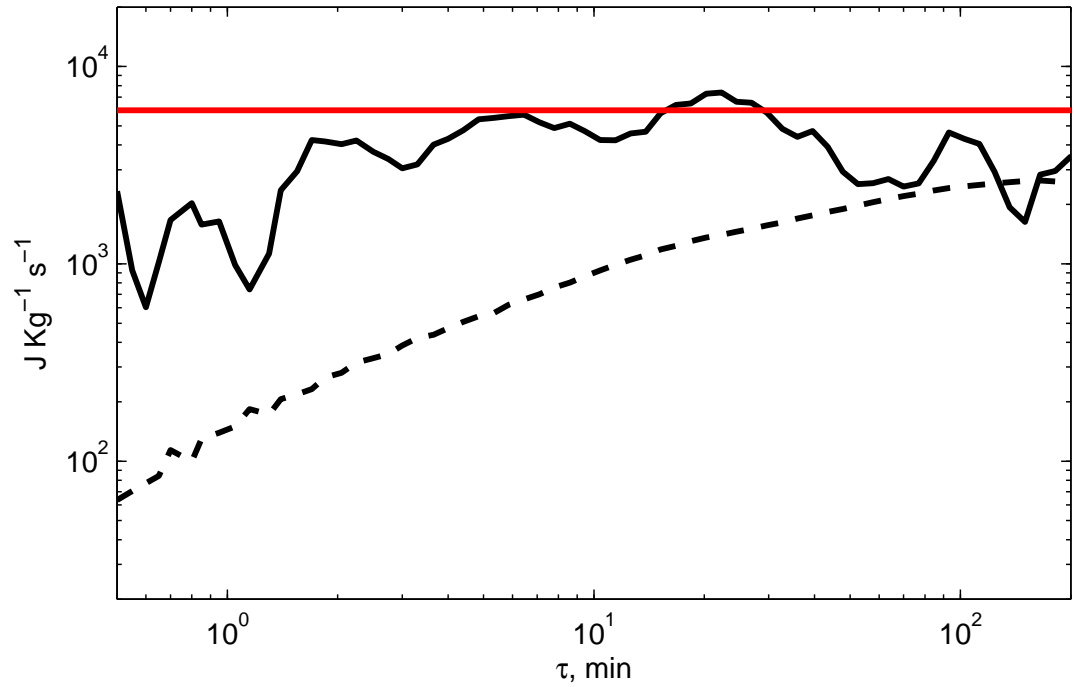}
\caption{The absolute value of the mean Yaglom term $|Y|$ (solid line), the cross-correlation term $M$ (dashed line) and the proton heating rate (red horizontal line) derived by means of equation (\ref{eq:PROTON}) for the fast stream detected by the WIND spacecraft.}\label{fig1}
\end{figure}

The energy cascade rate of MHD turbulence $\varepsilon$ is defined as $\varepsilon=(\varepsilon^+ + \varepsilon^-)/2$, so that,
\begin{equation}
\alpha Y + D + M = -\varepsilon. \label{eq:YAGt1}
\end{equation}
Here $Y=(Y^+ + Y^-)/2$ is the mean Yaglom term and $D=(D^+ + D^-)/2$ is the mean decay term. The data from the WIND satellite are provided in GSE, so that $\alpha=1$. The absolute value of the mean Yaglom term (solid line) and the cross-correlation term (dashed line) for the solar wind interval studied here  are presented in Figure \ref{fig1}. The mean energy dissipation rate derived from the Yaglom relation $\varepsilon =-Y$ is $\varepsilon\approx 5000 ~{\rm J~ Kg^{-1}~ s^{-1}}$ in qualitative agreement with previous studies \citep{SEA07,MSF08,OEA11}. The mean proton heating rate for the studied interval, derived by means of equation (\ref{eq:PROTON}), $\varepsilon_p=6050 ~{\rm J~ Kg^{-1}~ s^{-1}}$ is indicated by the red horizontal line in Figure 1. It has to be noted that the Yaglom's term is derived by averaging the third order mixed term that does not have a fixed sign, so that it requires much more data points for stable convergence with respect to the second order moment with fixed sign. Indeed, it has been showed that the stable convergence requires up to $10^5$ data points \citep{PEA09,SEA09}. If this condition is not fulfilled, then the Yaglom law is observed only in certain solar wind streams \citep{SEA07}. Our interval contains about $5\times 10^4$ points, thus giving quite stable results; they are found to be in agreement with other studies.

As it is well known the residual energy in the solar wind usually follows power law scaling in the inertial interval $\langle \Delta {\bf z}^+ \cdot \Delta {\bf z}^- \rangle \sim t^\gamma$ with $\gamma \approx 0.7$ (see, e.g., \cite{GCH12} and references therein). Thus, it is expected that in the inertial range the cross-correlation term in equations (\ref{eq:YAGr})-(\ref{eq:YAGt}) should behave as $M\sim t^\gamma$. On the other hand, at large scales correlations between fluctuations at different points are weakened, i.e., $\langle \Delta {\bf z}^+ \cdot \Delta {\bf z}^- \rangle \approx 2\langle {\bf z}^+ \cdot {\bf z}^- \rangle$, therefore, for very large time separations $M\rightarrow const$. This behaviour is clearly seen in Figure \ref{fig1}. 

It can be noted in Figure \ref{fig1} that at tens of minutes scales in the inertial range the Yaglom and the cross-correlation terms are of the same order of magnitude. The ratio of these two terms at the scale of 30 minutes $|Y|/M\approx 3.5$. We studied several tens of other intervals of fast solar wind streams and the mean value of this ratio at 30 min scales was $\langle |Y|/M \rangle \approx 2.5$. Usually, the mean energy cascade rate in the solar wind is determined as the average value of the dissipation rate derived via the relation $\varepsilon =-Y$ for time separations from 1 minute to 2 hours \citep{MSF08}. Since the cross-correlation term is always positive, neglecting $M$ can lead to an overestimation of the real energy dissipation rate by tens of percent. We also analysed quasi stationary intervals of slow solar wind streams and the analysis showed that the cross-correlation term in equation (\ref{eq:YAGt1}) is much less important for slow streams. The typical value of the ratio is $|Y|/M \approx 10-20$. These findings can explain some significant differences found between the energy dissipation rates obtained using equation (\ref{eq:PROTON}) and the estimation of the Yaglom law in fast solar wind streams observed by \cite{SEA09}. These authors found very good agreement between the proton heating rate and energy cascade rate for relatively cold, slow streams of the solar wind, whereas for hot, fast streams the Yaglom law provided significant overestimate of the cascade rate, compared to the value predicted by equation (\ref{eq:PROTON}). 

\begin{figure}
\epsscale{0.99}
\plotone{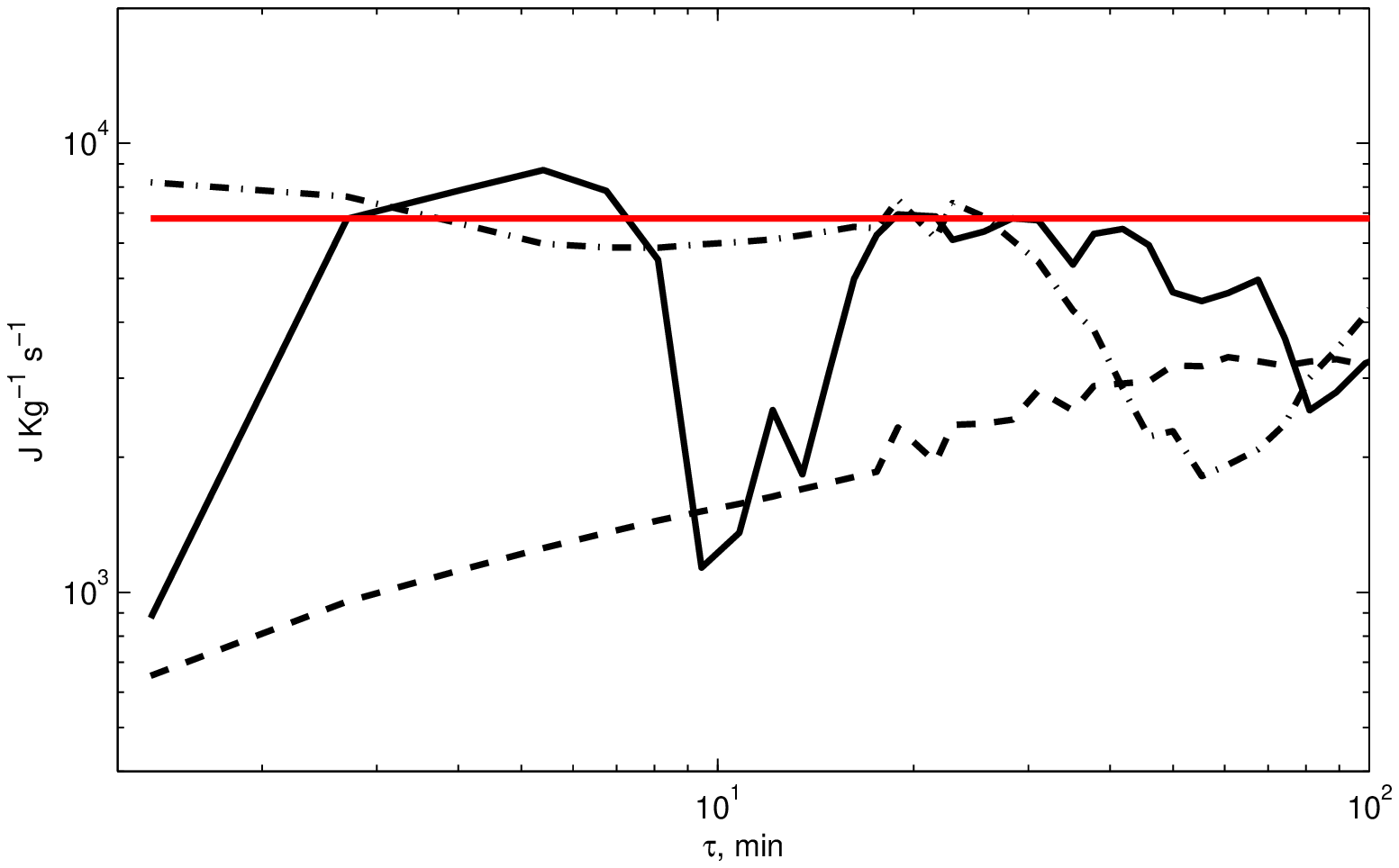}
\caption{The mean Yaglom term $Y$ (solid line), the cross-correlation $M$ term (dashed line) and the proton heating rate (red horizontal line) derived by means of equation (\ref{eq:PROTON}) from Helios 2 data at $0.87$ AU. The absolute value of the decay term $D$ at $0.65$ and $0.87$ AU is given by the dash-dotted line. See text for details.}\label{fig2}
\end{figure}

We also studied data from Helios 2 with 81 sec cadence collected from the same corotating stream at different heliospheric distances \citep{Bavassano82, MT90}. We used the data studied previously by \cite{ZM90}. The first interval is on 1976 day 76 when the distance from the sun was $0.65 ~{\rm AU}$. The start time of the second interval is 00:00 on 1976 day 50 and the end time is 22:04 on 1976 day 51, when the distance from the sun was $0.87 ~{\rm AU}$ \citep{MT90}. The data are provided in the RTN reference frame so that $\alpha =-1$ in equation (\ref{eq:YAGt1}). The mean Yaglom (solid line) and the cross-correlation (dashed line) terms at $0.87$ AU are presented in Figure \ref{fig2}. Their behaviour is similar to the one found at $1$ AU in the WIND data set, although the Yaglom term in Figure \ref{fig2} is less stable. This is due to the fact that the data set contains only about 3000 points. The absolute value of the mean decay term $D$ (which is actually negative due to decay of the turbulent energy during evolution) is given by the dash-dotted line in Figure \ref{fig2}. We used a linear approximation for the estimation of the derivative with respect to the heliospheric distance in the $D$ term of equation (\ref{eq:YAGt}). In particular, for any variable $Q$ we assumed $\partial_R[Q(R)]\approx [Q(0.87AU)-Q(0.65AU)]/[0.87AU-0.65AU]$. As can be seen, the Yaglom and the decay terms are pretty much comparable within all the range of time scales considered. In this contest one could conclude that the use of the Yaglom law without the decay term cannot give a reliable estimation of the cascade rate. However, it is worth stressing that using two intervals relevant to the same corotating stream, gives, in fact, an upper estimate of the decay term. Indeed, although the data belong to the same corotating stream, the time gap between the data sets is of several weeks, so that the assumption of stationary turbulence is questionable (indeed if the total turbulence energy changes not only because of nonlinear decay but also because of non-stationarity, both of those effects would influence the decay term); further, the rough two points approximation of the derivative with respect to the heliospheric distance $R$ can also contribute to the overestimation of the real value of $D$.  

\section{Conclusions}

In this letter we studied the Yaglom law for MHD turbulence in the expanding solar wind using the two scales expansion model by \citet{ZM90}. We derived the Yaglom law modified by two novel terms, which take into account the expansion effect. One of them is related to the energy decay by nonlinear interactions in MHD turbulence, while the second one is related to the non-zero cross-correlation of the Els\"asser fields. Using magnetic field and plasma data from WIND and Helios 2 spacecrafts, we show that for fast solar wind streams, at large time scales in the inertial range of solar wind turbulence, both the decay and the cross-correlation terms are comparable to the Yaglom's third order mixed moment, and, therefore, they can give a significant contribution in the assessment of the energy cascade rate. Thus, the disagreement between the proton heating rate estimate, obtained by using equation (\ref{eq:PROTON}), and the Yaglom law observed for fast streams in the solar wind can be ascribed to the fact that the extra terms in equation (\ref{eq:YAGr}) related to the expansion are usually neglected. 

\acknowledgments
S. P.'s research has been supported by ``Borsa Post-doc POR Calabria FSE 2007/2013 Asse IV Capitale Umano–Obiettivo Operativo M.2''. S. P. and V. C. acknowledge the Marie Curie Project FP7 PIRSES-2010-269297–``Turboplasmas''.


\end{document}